\documentclass[pra,twocolumn,showpacs]{revtex4}
\usepackage[dvips]{graphicx}
\input{epsf}
\usepackage{amsmath}
\usepackage{amssymb}
\usepackage[mathscr]{eucal}
\usepackage{bm}
\usepackage{theorem}

\newcommand{\beq}{\begin{equation}}
\newcommand{\eeq}{\end{equation}}
\newcommand{\beqa}{\begin{eqnarray}}
\newcommand{\eeqa}{\end{eqnarray}}
\newcommand{\beqan}{\begin{eqnarray*}}
\newcommand{\eeqan}{\end{eqnarray*}}

\newcommand{\tr}[1]{{\rm tr} \left( #1 \right) }
\newcommand{\ket}[1]{| #1 \rangle}
\newcommand{\bra}[1]{\langle #1 |}

\newcommand{\ox}{\otimes}
\newcommand{\op}{\oplus}

\newcommand{\proof}{\noindent {\bf Proof. }}
\newcommand{\qed}{\hfill $\Box$ \vskip 2ex}

\newtheorem{proposition}{Proposition}

{\theorembodyfont{\upshape}

}

\begin{document}

\title{Representing multiqubit unitary evolutions: spin coherences and infinitesimal coherences}
\author{Claudio Altafini}
\affiliation{SISSA-ISAS  \\
International School for Advanced Studies \\
via Beirut 2-4, 34014 Trieste, Italy }
\email{altafini@sissa.it}



\begin{abstract}
For the tensor of coherences parametrization of a multiqubit density operator, we provide an explicit formulation of the corresponding unitary dynamics at infinitesimal level.
The main advantage of this formalism (clearly reminiscent of the idea of ``coherences'' and ``coupling Hamiltonians'' of spin systems) is that the pattern of correlation between qubits and the pattern of infinitesimal correlation are highlighted simultaneously and can be used constructively for qubit manipulation.
For example, it allows to compute explicitly a Rodrigues' formula for the one-parameter orbits of nonlocal Hamiltonians.
The result is easily generalizable to orbits of Cartan subalgebras and allows to write the Cartan decomposition of unitary propagators as a linear action.

\end{abstract}

\maketitle 


\section{Introduction}

The easiest and most promising type of ``quantum network'' i.e., of collection of quantum systems to be manipulated individually or jointly for the purposes of quantum information processing, is by far composed of qubits i.e., of collections of two level systems.
For such systems, in \cite{Cla-qu-ent1} we investigated the use of a particular parametrization, called the tensor of coherences, obtained by the juxtaposition of an affine Bloch vector for each qubit, and of widespread use (with minor variations) under different names like cluster operators \cite{Mahler1}, Stokes tensor \cite{Jaeger2} or product of operator basis \cite{Ernst1,Havel1} in the literature on NMR spectroscopy. It is also closely related to multiparticle spacetime algebra \cite{Havel4} and to the recent work of Havel \cite{Havel3}.
Our tensor could be considered a particular parametrization of the ``nonsymmetric real density matrix'' of \cite{Havel3} especially suited to emphasize the Lie algebraic point of view of the equations of motion.

The scope of the present paper is to discuss how the differential equations describing unitary dynamics must be formulated in the tensor of coherences basis.
The idea that the unitary evolution of a qubit density matrix (pure or mixed) given by the Liouville-von Neumann equation becomes a linear vectorial ODE for the Bloch vector is generalized to multiqubit densities.
Mathematically, this could be thought of as ``passing to the adjoint representation'', its starting point being a formula for the decomposition of nonlocal commutators in terms of local commutators and anticommutators (see Appendix); practically it corresponds still to replacing a conjugation action on matrices with a linear action on the vector obtained by stacking the columns of the tensor.
In particular, when operations are local, a unitary transformation reduces to a multilinear action, i.e., a linear action on each piece of the tensor of coherences.
When instead nonlocal transformations are used, also their infinitesimal generators can be nonsplittable and multispin correlations are induced.
In this case the notation highlights which qubits are involved in correspondence of each nonlocal gate.
As a matter of fact, the major advantage of the formalism is that both the pattern of correlations (or ``coherences'' as they are called in the NMR spectroscopy literature) of the density tensor and the pattern of the couplings at infinitesimal level become very transparent as both are decomposed with respect to the same basis of observables.
In particular, they both show the same hierarchy of correlations (that originate from the affine structure of the tensors and of the corresponding Lie algebras of generators) which allows to keep track of all reduced dynamics and reduced densities in a natural way.
The idea of associating coherences to the degrees of freedom of qubits and of manipulating qubits through the corresponding Hamiltonians is common for example in the literature on spin systems in magnetic fields \cite{Havel1,Khaneja3,Rau2,Subrahmanyam1,Yung1}.
However, the principles apply to any network of qubits.
The price to pay is a larger dimension of the matrices representing the infinitesimal generators: while the size of the Hamiltonians grows as $ 2^n $ in the number $ n$ of qubits, in the adjoint representation it grows as $ 4^n = 2^{2n} $.

As an example of the insight gained into the dynamics of the system, we compute explicitly the integral flow of any nonlocal (constant) Hamiltonian, by means of a Rodrigues' formula \cite{Marsden1}, which shows that the exponential can be written as a sum of tensor products.
Since a Cartan subalgebra \cite{Khaneja2} contains only commuting vector fields, also the multiparameter orbit of a set of generators belonging to a Cartan subalgebra admits an explicit integration.
The Cartan decomposition becomes then a concatenation of local and nonlocal linear actions that can be expressed directly in terms of the infinitesimal generators, rather than of exponentials.
Such a decomposition has recently attracted considerable attention as a tool for constructing universal quantum gates which are optimal in the sense of time minimizers or complexity minimizers \cite{Khaneja3,ZhangJ1}.

A couple of other examples is discussed, mainly focused on the manipulation of qubits in presence of entanglement.
For example we show how to create entanglement at distance between qubits that are not directly coupled according to two different schemes, one in which the entanglement is distributed via an entangled ancilla, the other via a (always) separable ancilla as in \cite{Cubitt1}.

\section{Lie brackets and adjoint representation for spin $ \frac{1}{2} $ systems}
\label{sec:lie-br}
\subsection{One-spin}
\label{sec:one-spin}
Consider the rescaled Pauli matrices and identity matrix:
\[
\lambda_0 = \frac{1}{\sqrt{2}} \begin{bmatrix} 1 & 0 \\ 0 & 1 \end{bmatrix} \quad
\lambda_1 = \frac{1}{\sqrt{2}} \begin{bmatrix} 0 & 1 \\ 1 & 0 \end{bmatrix} 
\]
\[
\lambda_2 = \frac{1}{\sqrt{2}} \begin{bmatrix} 0 & -i \\ i & 0 \end{bmatrix} \quad
\lambda_3 = \frac{1}{\sqrt{2}} \begin{bmatrix} 1 & 0 \\ 0 & -1 \end{bmatrix}
\]
with the commutation relations 
\[
\begin{split}
[ \lambda_0 , \, \lambda_k ] = 0, \qquad
[ \lambda_1 , \, \lambda_2 ] = \sqrt{2} i \lambda_3, 
\\
[ \lambda_2 , \, \lambda_3 ] = \sqrt{2} i \lambda_1, \qquad 
[ \lambda_3 , \, \lambda_1 ] = \sqrt{2} i \lambda_2
\end{split}
\]
and the anticommutators
\beq
\begin{split}
& \{  \lambda_j , \, \lambda_k \}  = \sqrt{2} \delta_{jk}   \lambda_0  , \\
& \{  \lambda_j , \, \lambda_0 \}  =\{  \lambda_0 , \, \lambda_j \} = \sqrt{2} \lambda_j ,
\end{split} 
\label{eq:anticomm}
\eeq
$  j , \, k  \in \{ 1 , \, 2, \, 3\}$.
 The operator ``$ {\rm ad} $'' is defined as follows: $ {\rm ad}_{\lambda_j} \lambda_k = [ \lambda_j, \, \lambda_k ] = \sum_{l=0}^3 c_{jk}^l \lambda_l $ where operations involving the $0$ index only produce a null result: $ c_{0k}^l = c_{j0}^l = c_{jk}^0 =0 $.
Using the ``structure constants'' $ c_{jk}^l $ we obtain an ``adjoint basis'' associated to the $ \lambda_j $ matrices, given by the four $ 4\times 4 $ matrices $ {\rm ad}_{\lambda_0 }, \ldots, {\rm ad}_{\lambda_3 } $ of purely imaginary entries $ \left( {\rm ad}_{\lambda_j} \right)_{kl} = c_{jk}^l $:
\[
 {\rm ad}_{\lambda_0 } = 0_{4\times 4}, \qquad  
{\rm ad}_{\lambda_1 } =\sqrt{2} i \begin{bmatrix} 
0 & 0 & 0 & 0 \\
0 & 0 & 0 & 0 \\
0 & 0 & 0 & -1 \\
0 & 0 & 1 & 0
\end{bmatrix}
\]
\[
{\rm ad}_{\lambda_2 } =\sqrt{2} i \begin{bmatrix} 
0 & 0 & 0 & 0 \\
0 & 0 & 0 & 1 \\
0 & 0 & 0 & 0 \\
0 & -1 & 0 & 0
\end{bmatrix}
, \; \;
{\rm ad}_{\lambda_3 } = \sqrt{2} i \begin{bmatrix} 
0 & 0 & 0 & 0 \\
0 & 0 & -1 & 0 \\
0 & 1 & 0 & 0 \\
0 & 0 & 0 & 0
\end{bmatrix}
\]
The Pauli matrices are such that $ -i \lambda_1 $, $ -i \lambda_2 $ and $ -i \lambda_3 $ form a basis of $ \mathfrak{su}(2)$, while the $ -i {\rm ad}_{\lambda_j} $, $ j=1,\, 2,\, 3 $, form a basis of $ \mathfrak{so}(3) = {\rm ad}_{\mathfrak{su}(2) } $, the adjoint representation of $ \mathfrak{su}(2) $.
Concerning the ``antiadjoint'' operators $ {\rm aad}_{\lambda_j} $, $ j =0, \, 1, \, 2, \, 3 $, they can also be defined in the same fashion as the ${\rm ad}_{\lambda_j}  $, i.e., by means of $ 4\times 4 $ matrices obtained from $
{\rm aad}_{\lambda_j } \lambda_k = \{ \lambda_j , \, \lambda_k \} = \sum_{l=0}^3 s_{jk}^l \lambda_l $, $ j, \, k, \, l \in \{0, \, 1, \,2,\, 3\} $,
so that a linear representation of $ {\rm aad}_{\lambda_j } $ is given by $  \left( {\rm aad}_{\lambda_j }\right) _{kl} = s_{jk}^l $ with the $4\times 4 $ matrices $ {\rm aad}_{\lambda_j } $ easily computed from \eqref{eq:anticomm}:
\[
 {\rm aad}_{\lambda_0 } = \sqrt{2} \begin{bmatrix} 
1 & 0 & 0 & 0 \\
0 & 1 & 0 & 0 \\
0 & 0 & 1 & 0 \\
0 & 0 & 0 & 1
\end{bmatrix}, \; \;  
{\rm aad}_{\lambda_1 } =  \sqrt{2} \begin{bmatrix} 
0 & 1 & 0 & 0 \\
1 & 0 & 0 & 0 \\
0 & 0 & 0 & 0 \\
0 & 0 & 0 & 0
\end{bmatrix}
\]
\[
{\rm aad}_{\lambda_2 } = \sqrt{2} \begin{bmatrix} 
0 & 0 & 1 & 0 \\
0 & 0 & 0 & 0 \\
1 & 0 & 0 & 0 \\
0 & 0 & 0 & 0
\end{bmatrix}
, \; \; 
{\rm aad}_{\lambda_3 } =  \sqrt{2} \begin{bmatrix} 
0 & 0 & 0 & 1 \\
0 & 0 & 0 & 0 \\
0 & 0 & 0 & 0 \\
1 & 0 & 0 & 0
\end{bmatrix}
\]

\subsection{Two-spin}

Call $ \Lambda_{jk} = \lambda_j \otimes \lambda_k $, $  j, \,k \in \{ 0, \, 1 ,\, 2,\, 3\}  $.
Up to a constant, the $\Lambda_{jk} $ form the so-called {\em product operator basis}, see \cite{Ernst1}, and are subdivided into $0$ spin operators ($\Lambda_{00}$), $1$ spin operators ($\Lambda_{01}$, $\Lambda_{02}$, $\Lambda_{03}$, $\Lambda_{10}$, $\Lambda_{20}$, $\Lambda_{30}$) and $2$ spin operators ($\Lambda_{11}$, $\Lambda_{12}$, $\Lambda_{13}$, $\Lambda_{21}$, $\Lambda_{22}$, $\Lambda_{23}$, $\Lambda_{31}$, $ \Lambda_{32}$, $\Lambda_{33} $).
The set of $ -i \Lambda_{jk} $ $ j, \, k \in \{ 0, \, 1, \, 2,\, 3 \} $ contains a basis of the 9-dimensional tensor product Lie algebra $  \mathfrak{su}(2)\otimes  \mathfrak{su}(2)$ plus a basis of the 6-dimensional ``tensor sum'' Lie algebras $  \mathfrak{su}(2)\oplus  \mathfrak{su}(2)$ arising from the affine elements.
Just like $ \-i \lambda_0 \notin \mathfrak{su}(2)$, so $ -i \Lambda_{00} \notin  \mathfrak{su}(2)\otimes  \mathfrak{su}(2)$ and $ -i \Lambda_{00} \notin  \mathfrak{su}(2)\op  \mathfrak{su}(2)$. 
From \eqref{eq:app-Lieb1}:
\beq
\begin{split}
[ \Lambda_{jk} , \, \Lambda_{lm} ] & = [ \lambda_j \otimes \lambda_k , \, \lambda_l \otimes \lambda_m ] \\ 
& = {\rm ad}_{\Lambda_{jk}} \Lambda_{lm}  = {\rm ad}_{\lambda_j \otimes \lambda_k } \lambda_l \otimes \lambda_m  \\
& = \frac{1}{2} \left( [ \lambda_j , \, \lambda_l ] \otimes  \{  \lambda_k , \, \lambda_m \} + \{  \lambda_j , \, \lambda_l \} \otimes  [ \lambda_k, \, \lambda_m ] \right) \\
& =  \frac{1}{2} \left( 
{\rm ad}_{\lambda_j} \lambda_l \otimes {\rm aad}_{\lambda_k } \lambda_m 
+ {\rm aad}_{\lambda_j} \lambda_l \otimes {\rm ad}_{\lambda_k } \lambda_m 
\right) .
\end{split}
\label{eq:brack2spin}
\eeq
In terms of the adjoint representation, \eqref{eq:brack2spin} can be expressed as a 4-tensor, function of the two 2-tensors $ c_{jk}^l $ and $ s_{jk}^l $ as:
\beq
\begin{split}
{\rm ad}_{\Lambda_{jk}} & = {\rm ad}_{\lambda_j \otimes \lambda_k } \\
& =  \frac{1}{2} \left( {\rm ad}_{\lambda_j } \otimes  {\rm aad}_{\lambda_k } +  {\rm aad}_{\lambda_j } \otimes  {\rm ad}_{\lambda_k } \right) 
\end{split}
\label{eq:adj-rep-all}
\eeq
of elements
\beq
 \left( {\rm ad}_{\Lambda_{jk}} \right)_{lm}^{pq}  =\frac{1}{2} \left( c_{jl}^p \otimes s_{km}^q +  s_{jl}^p \otimes c_{km}^q \right) ,
\label{eq:adj-rep-elem}
\eeq
so that \eqref{eq:brack2spin} becomes:
\beq
\begin{split}
[ \Lambda_{jk} , \, \Lambda_{lm} ] & =\left( {\rm ad}_{\lambda_j \otimes \lambda_k } \right)_{lm}^{pq} \Lambda_{pq} \\
& = \frac{1}{2} \left( c_{jl}^p \otimes s_{km}^q +  s_{jl}^p \otimes c_{km}^q \right)  \Lambda_{pq}
\end{split}
\label{eq:brack2spin2}
\eeq
where we have used the summation convention over repeated indexes (in the range $ 0 \div 3$).
For $ j\neq 0 $ and $ k \neq 0 $, the $ -i {\rm ad}_{\Lambda_{jk}} $ of eq. \eqref{eq:adj-rep-all} form a basis of the adjoint representation of $ \mathfrak{su}(2) \ox \mathfrak{su}(2)  $, $  {\rm ad}_{ \mathfrak{su}(2) \ox\mathfrak{su}(2)  } = \mathfrak{so}(3) \ox \mathfrak{so}(3)   $. 
The remaining elements account for the affine structure i.e., for $ {\rm ad}_{ \mathfrak{su}(2) \op \mathfrak{su}(2)  } = \mathfrak{so}(3) \op \mathfrak{so}(3)$. As $ c_{jl}^p $ and $ s_{km}^q $ are $ 4 \times 4 $ matrices, the resulting Kronecker product $ {\rm ad}_{\Lambda_{jk}} $ is a $ 16\times 16 $ matrix.
However, it has a row and a column entirely composed of zeros in correspondence of $ \Lambda_{00} $ and, given $ \Lambda_{jk} $ with $ jk \neq 00 $, $\nexists $ $ \Lambda_{lm}  $ with $ (lm) \neq (00) $ such that $ {\rm ad}_{\Lambda_{jk}} \Lambda_{lm} = \Lambda_{00} $.
Furthermore, $ {\rm ad}_{\Lambda_{00}} $ being the trivial matrix of all zeros, it is not a basis element in the adjoint representation.
Also in the adjoint representation the index $0$ in a slot corresponds to trivial dynamics in the corresponding site. 
For example
\beq
\begin{split}
{\rm ad} _{\Lambda_{j0} } & =  \frac{1}{2} \left( {\rm ad}_{\lambda_j } \otimes  {\rm aad}_{\lambda_0 } +  {\rm aad}_{\lambda_j } \otimes 0 \right) \\
& =  \frac{1}{\sqrt{2}} {\rm ad}_{\lambda_j } \otimes I_4 .
\end{split}
\label{eq:adj-rep-j0}
\eeq

\subsection{$n$-spin} 
\label{sec:adj-n-splin}
In the $n$ spin case, $ \Lambda_{j_1\ldots j_n}=\lambda_{j_1} \ox \ldots \ox \lambda_{j_n} $, $ j_k \in \{ 0, \, 1, \, 2, \, 3 \} $, $ k \in \{ 1, \ldots , n \} $, are the basis elements. 
The Lie bracket $ [ \Lambda_{j_1\ldots j_n}, \,  \Lambda_{k_1\ldots k_n} ] $ can be computed according to the rule \eqref{n-fact-tens-Lieb}. 
For example, for $ n=3$ from \eqref{eq:app-Lieb2}:
\beq
\begin{split}
& [ \Lambda_{jkl} , \, \Lambda_{mpq} ]  = \frac{1}{4} \left( 
{\rm ad}_{\lambda_j} \lambda_m \otimes {\rm aad}_{\lambda_k } \lambda_p \ox  {\rm aad}_{\lambda_l } \lambda_q 
\right. \\
& \qquad  \qquad \qquad \left. 
+ {\rm aad}_{\lambda_j} \lambda_m \otimes {\rm ad}_{\lambda_k } \lambda_p \ox  {\rm aad}_{\lambda_l } \lambda_q 
\right. \\
& \qquad \qquad \qquad\left. 
+ {\rm aad}_{\lambda_j} \lambda_m \otimes {\rm aad}_{\lambda_k } \lambda_p \ox  {\rm ad}_{\lambda_l } \lambda_q 
\right. \\
& \qquad \qquad \qquad \left. 
+ {\rm ad}_{\lambda_j} \lambda_m \otimes {\rm ad}_{\lambda_k } \lambda_p \ox  {\rm ad}_{\lambda_l } \lambda_q 
\right) 
\\
& = \frac{1}{4} \left( 
{\rm ad}_{\lambda_j} \otimes {\rm aad}_{\lambda_k } \ox  {\rm aad}_{\lambda_l }
+ {\rm aad}_{\lambda_j} \otimes {\rm ad}_{\lambda_k }\ox  {\rm aad}_{\lambda_l } 
\right. \\
& \qquad \left.  
+ {\rm aad}_{\lambda_j} \otimes {\rm aad}_{\lambda_k } \ox  {\rm ad}_{\lambda_l } 
+ {\rm ad}_{\lambda_j} \otimes {\rm ad}_{\lambda_k }\ox  {\rm ad}_{\lambda_l } 
\right) ^{rst}_{mpq} \Lambda_{rst}
\\
& = \frac{1}{4} \left( 
c^r_{jm} \ox s^s_{kp} \ox s^t_{lq} 
+s^r_{jm} \ox c^s_{kp} \ox s^t_{lq} 
\right. \\ &\qquad \left.  
+s^r_{jm} \ox s^s_{kp} \ox c^t_{lq} 
+c^r_{jm} \ox c^s_{kp} \ox c^t_{lq} 
\right) ^{rst}_{mpq} \Lambda_{rst} \\
& =  \left( {\rm ad}_{\Lambda_{jkl}} \right) ^{rst}_{mpq} \Lambda_{rst}
\end{split}
\label{eq:brack3spin1}
\eeq
Remarkably, the building blocks needed for the $n$-qubit case are just the structure constants $ c_{jk}^l $ and $ s_{jk}^l $ computed above.
For $ n$ spins, the affine structure propagates itself throughout and determines a hierarchy of subalgebras of tensor product and tensor sum type.
The $ -i \Lambda_{j_1\ldots j_n}$, $ (j_1\ldots j_n) \neq (0 \ldots 0) $, form a joint basis of the Lie algebras $ \mathfrak{su}(2)^{\ox n }$, $  \mathfrak{su}(2) \op \mathfrak{su}(2)^{\ox (n-1) }$, $ \ldots $, $ \mathfrak{su}(2) \op \ldots \op \mathfrak{su}(2) $ (plus all factor permutations) and $ -i {\rm ad}_{\Lambda_{j_1\ldots j_n}} $, $ (j_1\ldots j_n) \neq (0 \ldots 0) $, a joint basis of $ {\rm ad}_{\mathfrak{su}(2)^{\ox n }}$, $  {\rm ad}_{\mathfrak{su}(2) \op \mathfrak{su}(2)^{\ox (n-1) }}$, $ \ldots $, $ {\rm ad}_{\mathfrak{su}(2) \op \ldots \op \mathfrak{su}(2) }$ (plus, again, all factor permutations). 
In both notations, the number and position of the indexes ``$ 0$'' uniquely determine which spins are involved into the $ -i {\rm ad}_{\Lambda_{j_1\ldots j_n}} $.

\section{Unitary evolution in terms of the tensor of coherences}
\label{sec:unit-ev}
For qubits, the same basis elements $  \Lambda_{j_1 \ldots j_n } $ that describe the infinitesimal generators can be used also for the density operators.
This is well-known in the literature on spin systems under the name of ``coherences'', \cite{Ernst1}, and can be formalized in terms of $ 4 \times 4 \times \ldots \times 4 $ tensors which we call tensors of coherences. See \cite{Cla-qu-ent1,Fano2,Jaeger2,Jakobczyk1,Mahler1,Verstraete1} for an overview.
The scope of this Section is to show how tensors of coherences and adjoint representation fit together in the description of unitary dynamics of multiqubit densities.

\subsection{Density operators and tensor of coherences}
This Section follows \cite{Cla-qu-ent1}.
The $ \Lambda_{j_1 \ldots j_n } $ form a complete orthonormal set for Hermitian matrices and can be used to obtain an affine tensorial representation of the density operator of $ n$ qubits: $
\rho=\varrho^{j_1 \ldots j_n} \Lambda_{j_1 \ldots j_n }$, $j_k \in \{ 0, \, 1, \, 2, \, 3 \} $, $ k \in \{ 1, \ldots , n \} $, with $ \varrho^{j_1 \ldots j_n} = \tr{\rho  \Lambda_{j_1 \ldots j_n } } $ the expectation value along the observables $ \Lambda_{j_1 \ldots j_n } $.
This representation has several advantages briefly recalled below:
\begin{itemize}
\item it captures all degrees of freedom of a density operator;
\item each term $ \varrho^{j_1\ldots j_n} $ in the tensor depends on a certain number of qubits: this is uniquely determined by the number of nonzero indexes in the sequence $ j_1 \ldots j_n $.
The pattern of nonzero indexes also identifies which qubits are involved, formalizing the idea of coherences of widespread use for spin systems. 
\item all correlations of all orders and all reduced densities are already contained in the tensor: tracing out a qubit means collapsing the corresponding index to $0$ and rescaling everything by $ \sqrt{2}$.
For example, if $ \rho_{A_2 \ldots A_n} = {\rm tr}_{A_1} \left( \rho \right) = \varrho^{j_2\ldots j_n} \Lambda_{j_2 \ldots j_n } $ then $  \varrho^{j_2\ldots j_n} = \sqrt{2}  \varrho^{0 j_2\ldots j_n} $;
\item Since 
\beq
\tr{\Lambda_{jk}\Lambda_{lm} }= \delta_{jl}\delta_{km}, 
\label{eq:tr-tens}
\eeq
$ j, \, k, \, l, \, m \in \{ 0, \, 1, \, 2, \, 3 \} $, the degree of mixing becomes the Euclidean norm of $\varrho^{j_1 \ldots j_n} $:
\beq
\tr{\rho} = \sum_{j_1 \ldots j_n =0 }^3 \left( \varrho^{j_1 \ldots j_n} \right)^2 
\label{eq:tr-sq}
\eeq
and hence, since $ \varrho^{0\ldots 0} = \left( 1/\sqrt{2} \right)^{n} $, for $ (j_1 \ldots j_n) \neq (0 \ldots 0) $ the tensor $ \varrho^{j_1 \ldots j_n} \in \mathbb{S}^{4^n-1 }_{r}\in \mathbb{R}^{4^n-1} $ with $ 0 \leqslant r \leqslant \sqrt{1 -\left( \varrho^{0\ldots 0}\right)^{2}} =\sqrt{1 -\left( 1/2 \right)^{n}} $;
\item complete mixing corresponds to $ r=0 $ (i.e., to the null tensor except for the affine constant $ \varrho^{0\ldots 0}$): 
\item pure states correspond to $ r = \sqrt{1 -\left( 1/2 \right)^{n}} $;
\item uncorrelation corresponds to $ \varrho^{j_1 \ldots j_n} = \varrho^{j_1}_{A_1} \varrho^{j_2}_{A_2} \ldots \varrho^{j_n}_{A_n} $,
where $ \varrho_{A_1} ^{j_1} = (\sqrt{2})^{ n-1} \varrho^{j_1 0 \ldots 0 } $ is the 4-vector of the reduced density $ \rho_{A_1} = {\rm tr}_{A_2\ldots A_n} \left( \rho\right) $ and so on;
\item partial transposition of a qubit becomes a change of sign in the terms having index $2$ in the corresponding slot. For example
\beq
\begin{split}
 \qquad \rho^{T_{A_1}} = & \; \; 
\varrho^{0 j_2 \ldots j_n} \Lambda_{0 j_2 \ldots j_n } 
+ \varrho^{1 j_2 \ldots j_n} \Lambda_{1 j_2 \ldots j_n } 
\\ & 
- \varrho^{2 j_2 \ldots j_n} \Lambda_{2 j_2 \ldots j_n } 
+ \varrho^{3 j_2 \ldots j_n} \Lambda_{3 j_2 \ldots j_n }
\end{split}
\label{eq:PT1}
\eeq
and so on;
\item checking bipartite entanglement can be done by the simple test \eqref{eq:PT1}.
\end{itemize}

\subsection{Liouville-von Neumann equation}
\label{sec:Liouville}
The Liouville-von Neumann equation for the $n$-qubits density $ \rho $ is:
\beq
\dot \rho = - i [ H , \, \rho ] = -i {\rm ad} _H ( \rho) 
\label{eq:Liouville1}
\eeq
where $ H = H^\dagger $ is the Hamiltonian of the system.
From Section~\ref{sec:lie-br}, we have that $ H = h^{j_1\ldots j_n} \Lambda_{j_1\ldots j_n}$, $ j_k \in \{0, \, 1, \,2,\, 3\}$, $ k \in \{ 1, \ldots , n \}$.

If we have two qubits then, in terms of the tensor of coherences, eq. \eqref{eq:Liouville1} corresponds to:
\beq
\begin{split}
\dot{\varrho}^{pq} & = - i  h^{jk} \left( {\rm ad}_{\Lambda_{jk}} \right) _{lm}^{pq} \varrho^{lm} \\
& = - \frac{i  h^{jk}}{2} \left(c_{jl}^p \otimes s_{km}^q +  s_{jl}^p \otimes c_{km}^q\right)  \varrho^{lm}.
\end{split}
\label{eq:Liouv-vectcoh1}
\eeq
In order to show \eqref{eq:Liouv-vectcoh1}, derive $ \varrho^{pq} = \tr{\rho \Lambda_{pq} } $ and use \eqref{eq:adj-rep-elem} and \eqref{eq:tr-tens}:
\beqan
\dot{\varrho}^{pq} & = & \tr{\dot{\rho} \Lambda_{pq} } 
= \tr{ -i [ H, \, \rho] \Lambda_{pq} } \\
& = &  \tr{ -i  h^{jk} [ \Lambda_{jk}, \, \Lambda_{lm} ] \varrho^{lm} \Lambda_{pq} } \\
& = & \tr{ -\frac{i}{2}  h^{jk} \left(c_{jl}^r \otimes s_{km}^s +  s_{jl}^r \otimes c_{km}^s\right)\Lambda_{rs}  \varrho^{lm} \Lambda_{pq} } \\
& = &  - \frac{i}{2} h^{jk} \left(c_{jl}^r \otimes s_{km}^s +  s_{jl}^r \otimes c_{km}^s\right)\varrho^{lm} \tr{\Lambda_{rs}  \Lambda_{pq} } \\
& =&   - \frac{i}{2}  h^{jk}\left(c_{jl}^r \otimes s_{km}^s +  s_{jl}^r \otimes c_{km}^s\right)\varrho^{lm} \delta_{rp}\delta_{sq} \\
& = & - \frac{i}{2} h^{jk}\left(c_{jl}^p \otimes s_{km}^q +  s_{jl}^p \otimes c_{km}^q\right)\varrho^{lm}
\eeqan
The component of the Hamiltonian along $ \Lambda_{00} $ is irrelevant: even if $ h^{00} \neq 0 $ it must be $ -i h^{00} {\rm ad}_{\Lambda_{00}}=0 $.
The meaning is similar to the single spin case: global phases are neglected in \eqref{eq:Liouville1} and \eqref{eq:Liouv-vectcoh1}.

Since \eqref{eq:Liouv-vectcoh1} is a linear system, if $ h^{jk} $ are constant the integration can be carried out explicitly: 
\beq
\varrho^{pq}(t)  =\left( e^{ - i  t h^{jk} {\rm ad}_{\Lambda_{jk}}} \right) _{lm}^{pq} \varrho^{lm} (0) .
\label{eq:liouv-integr}
\eeq
Notice that when 2-spin coherences are lacking, $ h^{jk}=0$ $ \forall \; j\neq 0 $ and $ k\neq 0$, i.e., when only LOCC operations are performed, the exponential in \eqref{eq:liouv-integr} splits.
In fact, $ [\Lambda_{j0}, \, \Lambda_{0k} ] =0 $ and therefore the infinitesimal generators $ \Lambda_{j0} $ and $ \Lambda_{0k} $ can be ``reduced'' as well. 
From \eqref{eq:adj-rep-j0}, the unitary propagator in eq. \eqref{eq:liouv-integr} becomes:
\[
\begin{split}
& e^{ - i  t \left( h^{j0} {\rm ad}_{\Lambda_{j0}} +  h^{0k} {\rm ad}_{\Lambda_{0k}} \right) } = \left(  e^{ - i  t  h^{j0} {\rm ad}_{\Lambda_{j0}} }\right)   \left(  e^{ - i  t  h^{0k} {\rm ad}_{\Lambda_{0k}} }\right)   \\
& = \left( \left(  e^{ - i  t \frac{h^{j0}}{\sqrt{2}} {\rm ad}_{\lambda_{j}}} \right) \ox I_4 \right) \left( I_4 \ox \left(  e^{ - i  t \frac{h^{0k}}{\sqrt{2}} {\rm ad}_{\lambda_{k}}} \right) \right) 
\end{split}
\]
where the factor $ \frac{1}{\sqrt{2}} $ comes from \eqref{eq:adj-rep-j0}.
Therefore
\beq
 \left(  e^{ - i  t \frac{h^{j0}}{\sqrt{2}} {\rm ad}_{\lambda_{j}}} \right) \ox
\left(  e^{ - i  t \frac{h^{0k}}{\sqrt{2}} {\rm ad}_{\lambda_{k}}} \right) 
\in \begin{bmatrix} 1 & 0 \\
0 & SO(3) 
\end{bmatrix} 
\ox 
\begin{bmatrix} 1 & 0 \\
0 & SO(3) 
\end{bmatrix} 
\label{eq:local-evol}
\eeq
which allows the state to evolve on at most a 6-parameter orbit sitting inside the 15-dimensional affine sphere $ \mathbb{S}^{15}_r $, with $r$ identified by \eqref{eq:tr-sq}.
If $\rho(0) $ is separable then under \eqref{eq:local-evol} so is $ \rho(t) $ for all $ t$, the 6-dimensional manifold contains all the separable states.
When instead the Hamiltonian has $ h^{jk}\neq 0 $ for $ j\neq 0 $ and $ k\neq 0 $, the evolution of the two qubits becomes coupled.

Similarly to the 2-qubit case, if we have $n$ qubits we obtain:
\[
\dot{\varrho}^{p_1\ldots p_n} = - i  h^{j_1\ldots j_n} \left( {\rm ad}_{\Lambda_{j_1\ldots j_n}} \right) _{k_1\ldots k_n}^{p_1\ldots p_n} \varrho^{k_1\ldots k_n} ,
\]
where $  {\rm ad}_{\Lambda_{j_1\ldots j_n}}  $ is computed as in Section~\ref{sec:adj-n-splin}.

\section{Integral flow of nonlocal Hamiltonians}
We first restrict to 2 qubits, although all arguments generalize to $n$ qubits.
First we give an explicit formula for the integral of each ``elementary'' generator $ \Lambda_{jk} $.
From Section~\ref{sec:one-spin}, we have that $
{\rm ad}_{\lambda_j}  {\rm aad}_{\lambda_j} = {\rm aad}_{\lambda_j}  {\rm ad}_{\lambda_j} =0 $.
This implies that the series expansion $ {\rm exp} \left( -i t {\rm ad}_{\Lambda_{jk} } \right) = \sum_{p=0}^\infty \frac{( -i t)^p}{p!}  {\rm ad}_{\Lambda_{jk}}^p $ has a particularly simple expression, since for all $ p$ 
\[
 {\rm ad}_{\Lambda_{jk}}^p = \frac{1}{2^p} \left(  {\rm ad}_{\lambda_{j}}^p \ox {\rm aad}_{\lambda_{k}}^p +  {\rm aad}_{\lambda_{j}}^p \ox {\rm ad}_{\lambda_{k}}^p \right) .
\]
The powers of $ {\rm ad}_{\lambda_j}$ and $  {\rm aad}_{\lambda_j} $ are easily computed since $ {\rm ad}_{\lambda_j}^2 $ and $ {\rm aad}_{\lambda_j}^2 $ are diagonal and ``complementary'':
\begin{itemize}
\item if $ j=1 $, $  {\rm ad}_{\lambda_1}^2 = 2 ( \delta_{33} + \delta_{44} ) $, $ {\rm aad}_{\lambda_1}^2 = 2 ( \delta_{11} + \delta_{22} ) $;
\item if $ j=2 $, $  {\rm ad}_{\lambda_2}^2 = 2 ( \delta_{22} + \delta_{44} ) $, $ {\rm aad}_{\lambda_2}^2 = 2 ( \delta_{11} + \delta_{33} ) $;
\item if $ j=3 $, $  {\rm ad}_{\lambda_3}^2 = 2 ( \delta_{22} + \delta_{33} ) $, $ {\rm aad}_{\lambda_3}^2 = 2 ( \delta_{11} + \delta_{44} ) $;
\end{itemize}
so that $ {\rm ad}_{\lambda_j}^2 + {\rm aad}_{\lambda_j}^2 = 2 I_4$.
Cubic powers instead are $  {\rm ad}_{\lambda_j}^3= 2 {\rm ad}_{\lambda_j} $ and $  {\rm aad}_{\lambda_j}^3= 2 {\rm aad}_{\lambda_j} $, hence $   {\rm ad}_{\Lambda_{jk}}^3=  {\rm ad}_{\Lambda_{jk}} $.
We can therefore explicitly write down the sum of the series as 
\[
\begin{split}
{\rm exp} \left( -i t {\rm ad}_{\Lambda_{jk} } \right) & = I_4 \ox I_4 -i \left( t - \frac{t^3}{3!} + \frac{t^5}{5!} - \ldots \right) {\rm ad}_{\Lambda_{jk} } \\
&  + \left( - \frac{t^2}{2!} + \frac{t^4}{4!} - \ldots \right)  {\rm ad}_{\Lambda_{jk} } ^2 
\end{split}
\]
or, adding and subtracting $  {\rm ad}_{\Lambda_{jk} } ^2 $,
\beq
{\rm exp} \left( -i t {\rm ad}_{\Lambda_{jk} } \right) =  I_4 \ox I_4 - i \sin (t)  {\rm ad}_{\Lambda_{jk} } - (1 - \cos (t) ) {\rm ad}_{\Lambda_{jk} } ^2 
\label{eq:int-of-mot1}
\eeq
where the extra terms added are needed because the zero order terms do not match: $ I_4 \ox I_4 \neq {\rm ad}_{\Lambda_{jk} } ^2 $.
Notice that a formula equivalent to \eqref{eq:int-of-mot1} was used for the same purposes as ours in \cite{Havel3}.
Both are tensored versions of the Rodrigues' formula for rotations, see \cite{Marsden2}, p. 291 or \cite{Murray1}, p. 28.
The splitting is into skew-symmetric ($ -i  {\rm ad}_{\Lambda_{jk} } $) and symmetric part ($  I_4 \ox I_4 $ and $  {\rm ad}_{\Lambda_{jk} }^2 $) of the flow \footnote{Even the sign difference with respect to the standard $SO(3)$ formula is due to the fact that here the skew-symmetric generator is $ -i  {\rm ad}_{\Lambda_{jk} } $.}.
Notice that both $ -i {\rm ad}_{\Lambda_{jk} } $ and $ {\rm ad}_{\Lambda_{jk} } ^2 $ are tensor products of matrices.
The nonlocality of the Hamiltonian of \eqref{eq:int-of-mot1} reflects in the fact that we do not obtain a ``single'' tensor product but rather a sum \footnote{This does not mean that we have separable superoperators \cite{Bennett2} however, since unitary operators yield ``pure'' quantum operations \cite{Nielsen3}.}.
Clearly the overall evolution of \eqref{eq:int-of-mot1} is orthogonal.
However, the single pieces do not describe rotations, neither locally nor globally.

The same argument can be repeated for any number of qubits. 
For example for 3 qubits we have $  {\rm exp} \left( -it {\rm ad}_{\Lambda_{jkl}} \right) = \sum_{p=0} ^\infty \frac{(-i t )^p}{p!}  {\rm ad}_{\Lambda_{jkl}}^p  $, with
\beq
\begin{split}
 {\rm ad}_{\Lambda_{jkl}}^p & \! \! \! \! = \frac{1}{4^p} \!  \left(  
{\rm ad}_{\lambda_j} ^p  \! 
\ox {\rm aad}_{\lambda_k} ^p  \!
\ox {\rm aad}_{\lambda_l} ^p  \! \!
+   {\rm aad}_{\lambda_j} ^p  \!
\ox {\rm ad}_{\lambda_k} ^p  \! 
\ox {\rm aad}_{\lambda_l} ^p 
\right. \\ & \left. 
+   {\rm aad}_{\lambda_j} ^p 
\ox {\rm aad}_{\lambda_k} ^p 
\ox {\rm ad}_{\lambda_l} ^p 
+   {\rm ad}_{\lambda_j} ^p 
\ox {\rm ad}_{\lambda_k} ^p 
\ox {\rm ad}_{\lambda_l} ^p 
\right)
\end{split}
\label{eq:adLjkl^p}
\eeq
where now $  {\rm ad}_{\Lambda_{jkl}}^3 = \frac{1}{2}  {\rm ad}_{\Lambda_{jkl}} $.
The sum of the series is then
\beq
\begin{split}
{\rm exp} \left( -i t {\rm ad}_{\Lambda_{jkl} } \right)  = & \,  I_4^{ \ox 3}  - i \sqrt{2}  \sin (\frac{t}{\sqrt{2}})  {\rm ad}_{\Lambda_{jkl} } \\ 
& - 2 \left(1 - \cos (\frac{t}{\sqrt{2}}) \right) {\rm ad}_{\Lambda_{jkl} } ^2. 
\end{split}
\label{eq:int-of-mot2}
\eeq

So far we have only considered a single ``coordinate direction'' ($\Lambda_{jk} $ for the 2-qubit case).
The formul{\ae} however extend in a straightforward manner to linear combinations of commuting generators, even depending on more than one parameter.
A maximal orbit of integrable flow is obtained obviously in correspondence of a Cartan subalgebra \cite{Khaneja2,ZhangJ1}, i.e., a maximal commuting subalgebra in the Lie algebra of nonlocal operations of the system.
For the 2-qubit case, let us concentrate on the ``nonlocal subalgebra'' $ {\rm ad}_{\mathfrak{su}(2) \ox \mathfrak{su}(2) }=\mathfrak{so}(3) \ox \mathfrak{so}(3)  $. 
A Cartan subalgebra is for example given by $ \mathfrak{h} = {\rm span} \{ -i  {\rm ad}_{\Lambda_{11}},\,  -i  {\rm ad}_{\Lambda_{22}}, \,  -i {\rm ad}_{\Lambda_{33}} \} $ (or by $ {\rm span} \{  -i {\rm ad}_{\Lambda_{12}},\,  -i {\rm ad}_{\Lambda_{21}}, \,  -i {\rm ad}_{\Lambda_{33}} \} $, etc.).
The 3-parameter orbits of such subalgebras are integrable as can be seen by the splitting of the exponential 
\beq
\begin{split}
& {\rm exp} \left( -i \left( \alpha^{11} {\rm ad}_{\Lambda_{11} } + \alpha^{22} {\rm ad}_{\Lambda_{22} } + \alpha^{33} {\rm ad}_{\Lambda_{33} } \right) \right)  = \\
& = {\rm exp} \left( -i \alpha^{11} {\rm ad}_{\Lambda_{11} } \right) {\rm exp} \left( -i \alpha^{22} {\rm ad}_{\Lambda_{22} }  \right) {\rm exp} \left( -i \alpha^{33} {\rm ad}_{\Lambda_{33} } \right) 
\end{split}
\label{eq:int-mot-Cartan}
\eeq
for real $ \alpha^{jj} $. 
The ``marginal'' subalgebra of local operations $ \mathfrak{so}(3) \op \mathfrak{so}(3)  $ does not commute with the Cartan subalgebras.
It is known \cite{Khaneja2} that $ [ \mathfrak{so}(3) \op \mathfrak{so}(3) , \mathfrak{h} ] $ generates the entire 15-dimensional Lie algebra $  \mathfrak{so}(3) \op \mathfrak{so}(3) \cup\mathfrak{so}(3) \ox \mathfrak{so}(3) $ and that ``exponentiating'' this splitting we get the Cartan decomposition of the corresponding Lie group.
With our formalism, such conjugation action becomes a linear action, obtained by the concatenation of a bilocal exponential as \eqref{eq:local-evol} and of \eqref{eq:int-mot-Cartan}.
In other words, any unitary operation acting on a 2-qubit density can be written for the tensor of coherences as the concatenation:
\[
\begin{split}
& \left(  e^{ - i \frac{\alpha^{j0}}{\sqrt{2}} {\rm ad}_{\lambda_{j}}} \right) \ox
\left(  e^{ - i  \frac{\alpha^{0k}}{\sqrt{2}} {\rm ad}_{\lambda_{k}}} \right)  {\rm exp} \left( -i \alpha^{11} {\rm ad}_{\Lambda_{11} } \right)
\\ & \quad \cdot {\rm exp} \left( -i \alpha^{22} {\rm ad}_{\Lambda_{22} }  \right) {\rm exp} \left( -i \alpha^{33} {\rm ad}_{\Lambda_{33} } \right) 
\end{split}
\]
for real $ \alpha^{jk}$.
Each exponential can be replaced by the corresponding sum of tensors (given by \eqref{eq:int-of-mot1} for the nonlocal pieces and by $  {\rm exp} \left( -i t {\rm ad}_{\lambda_{j} } \right) =  I_4 - \frac{i}{\sqrt{2}} \sin (\sqrt{2} t)  {\rm ad}_{\lambda_{j} } - \frac{(1 - \cos (\sqrt{2} t) )}{2}  {\rm ad}_{\lambda_{j} } ^2 $ for the one-parameter orbit of a single qubit).

\section{Examples}
In Example~\ref{sec:ex:C-NOT} it is shown how to express in term of the tensor of coherences the discrete unitary propagator corresponding to a standard 2-qubit gate, the {\rm C-NOT} gate.
In the three qubits of Example~\ref{sec:ex:3qub-ent-at-dist}, entangling between two ``distant'' qubits is achieved through indirect coupling by means of an entangled ancilla.
In Example~\ref{sec:ex:3qub-ent-at-distII}, instead, for the same purposes the scheme of \cite{Cubitt1} is used, in which the ancilla remains separable for all times.

\subsection{{\rm C-NOT} gate}
\label{sec:ex:C-NOT}
It is well-known that elementary gaits of a quantum computer being discrete unitary operations, they can written in terms of the corresponding infinitesimal Hamiltonians. In particular, in the literature on quantum information processing by means of NMR spectroscopy \cite{Havel1} this was done in terms of the product of operator bases of which our formalism is just a variation.
For example, in correspondence of the computational basis of two qubits $ \ket{00}$, $ \ket{01}$, $ \ket{10}$, $\ket{11}$, the Hamiltonian of the C-NOT gate 
\[
U_{\rm C-NOT} =  \begin{bmatrix} 
1 & 0 & 0 & 0 \\
0 & 0 & 0 & 1 \\
0 & 0 & 1 & 0 \\
0 & 1 & 0 & 0 
\end{bmatrix}
\]
is given by \cite{Havel1} 
\[
H_{\rm C-NOT} = \frac{\pi}{2} \begin{bmatrix} 
0 & 0 & 0 & 0 \\
0 & 1 & 0 & -1 \\
0 & 0 & 0 & 0 \\
0 & -1 & 0 & 1 
\end{bmatrix}.
\]
In terms of the $ \Lambda_{jk} $, this is $ H_{\rm C-NOT} = \frac{\pi}{2} \left( \Lambda_{00} -  \Lambda_{03} - \Lambda_{10} +  \Lambda_{13} \right) $, and therefore for $ \varrho^{jk} $ we have the orthogonal matrix
\[
R_{\rm C-NOT} = e^{- i \frac{\pi}{2} \left(  {\rm ad}_{\Lambda_{00}}- {\rm ad}_{\Lambda_{03}} - {\rm ad}_{\Lambda_{10}} + {\rm ad}_{\Lambda_{13}} \right) },
\]
which computed by means of \eqref{eq:adj-rep-all} yields 
\[ \! \! 
R_{\rm C-NOT} =  
\left[ \begin{array}{cccccccccccccccc}
1 & 0 & 0 & 0 & 0 & 0 & 0 & 0 & 0 & 0 & 0 & 0 & 0 & 0 & 0 & 0 \cr
0 & 0 & 0 & 0 & 0 & 1 & 0 & 0 & 0 & 0 & 0 & 0 & 0 & 0 & 0 & 0 \cr 
0 & 0 & 0 & 0 & 0 & 0 & 1 & 0 & 0 & 0 & 0 & 0 & 0 & 0 & 0 & 0 \cr 
0 & 0 & 0 & 1 & 0 & 0 & 0 & 0 & 0 & 0 & 0 & 0 & 0 & 0 & 0 & 0 \cr 
0 & 0 & 0 & 0 & 1 & 0 & 0 & 0 & 0 & 0 & 0 & 0 & 0 & 0 & 0 & 0 \cr 
0 & 1 & 0 & 0 & 0 & 0 & 0 & 0 & 0 & 0 & 0 & 0 & 0 & 0 & 0 & 0 \cr 
0 & 0 & 1 & 0 & 0 & 0 & 0 & 0 & 0 & 0 & 0 & 0 & 0 & 0 & 0 & 0 \cr 
0 & 0 & 0 & 0 & 0 & 0 & 0 & 1 & 0 & 0 & 0 & 0 & 0 & 0 & 0 & 0 \cr 
0 & 0 & 0 & 0 & 0 & 0 & 0 & 0 & 0 & 0 & 0 & 1 & 0 & 0 & 0 & 0 \cr 
0 & 0 & 0 & 0 & 0 & 0 & 0 & 0 & 0 & 0 & 0 & 0 & 0 & 0 & 1 & 0 \cr 
0 & 0 & 0 & 0 & 0 & 0 & 0 & 0 & 0 & 0 & 0 & 0 & 0 & -1 & 0 & 0 \cr 
0 & 0 & 0 & 0 & 0 & 0 & 0 & 0 & 1 & 0 & 0 & 0 & 0 & 0 & 0 & 0 \cr 
0 & 0 & 0 & 0 & 0 & 0 & 0 & 0 & 0 & 0 & 0 & 0 & 0 & 0 & 0 & 1 \cr 
0 & 0 & 0 & 0 & 0 & 0 & 0 & 0 & 0 & 0 & -1 & 0 & 0 & 0 & 0 & 0 \cr 
0 & 0 & 0 & 0 & 0 & 0 & 0 & 0 & 0 & 1 & 0 & 0 & 0 & 0 & 0 & 0 \cr 
0 & 0 & 0 & 0 & 0 & 0 & 0 & 0 & 0 & 0 & 0 & 0 & 1 & 0 & 0 & 0 
\end{array} \right].
\]
If we are given the 4 computational basis states
\[
\begin{split}
\ket{00} & \longleftrightarrow \varrho^{jk} = \left\{ \begin{array}{l} 1/2, 0, 0, 1/2, 0, 0, 0, 0,  \\
 0, 0, 0, 0, 1/2, 0, 0, 1/2 \end{array} \right\},  \\
\ket{01} & \longleftrightarrow \varrho^{jk} = \left\{\begin{array}{l} 1/2, 0, 0, -1/2, 0, 0, 0, 0, \\
 0, 0, 0, 0, 1/2, 0, 0, -1/2 \end{array} \right\},  \\
\ket{01} & \longleftrightarrow \varrho^{jk} = \left\{ \begin{array}{l} 1/2, 0, 0, 1/2, 0, 0, 0, 0, \\
0, 0, 0, 0, -1/2, 0, 0, -1/2 \end{array} \right\},  \\
\ket{11} & \longleftrightarrow \varrho^{jk} =  \left\{ \begin{array}{l}  1/2, 0, 0, -1/2, 0, 0, 0, 0,\\
 0, 0, 0, 0, -1/2, 0, 0, 1/2   \end{array} \right\} ,
\end{split}
\]
it is straightforward to check that $ R_{\rm C-NOT}  $ behaves as a C-NOT gate with the second qubit acting as control qubit.
Notice that $ H_{\rm C-NOT}$ is not traceless, hence we have an Hamiltonian with $ h^{00} \neq 0 $. 
As mentioned above, this is irrelevant because $ {\rm ad}_{\Lambda_{00} } =0 $, i.e., in the adjoint representation one always obtains the corresponding traceless Hamiltonian.

The structure of the basis used indicates that $ H_{\rm C-NOT} $ is a non-local operation since it contains $ \Lambda_{13} $ (and the splitting into basis elements is obviously unique).
While it leaves unentangled the computational basis elements, the same is not true in general for any state. 

Comparing $ U_{\rm C-NOT} $ and $ R_{\rm C-NOT}$, the price to pay in order to use the tensor of coherences parametrization is a larger dimension of the operator involved. On the other hand, the matrices are normally sparse and the formalism allows to perform the same operation also on mixed states.

\subsection{Three-qubit: entangling at distance (I)}
\label{sec:ex:3qub-ent-at-dist}
\begin{figure}[ht]
\begin{center}
 \includegraphics[width=9cm]{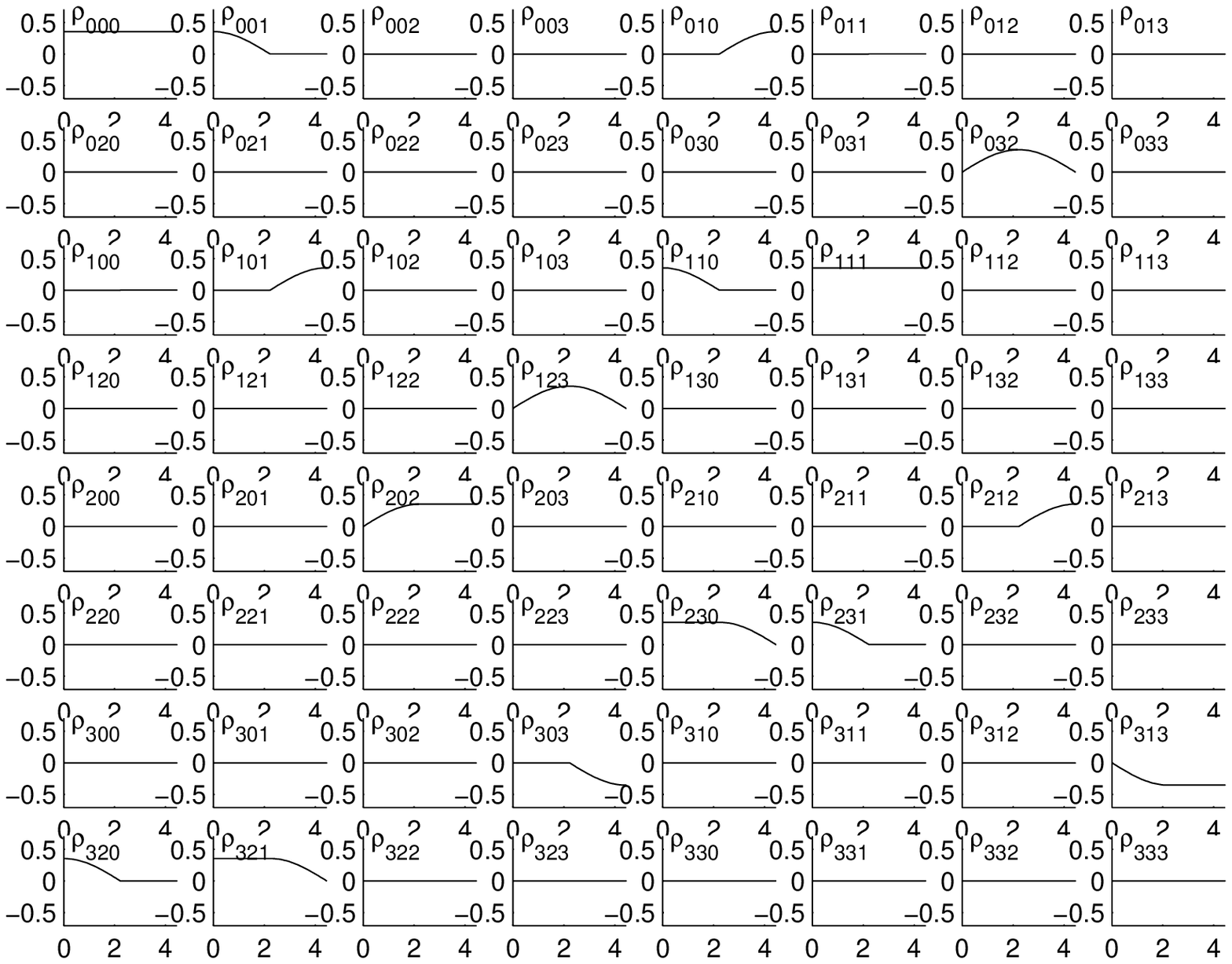}
 \caption{The 64 components of $ \rho$ versus $ t $.}
\label{fig:ex3a}
\end{center}
\end{figure}

\begin{figure}[ht]
\begin{center}
 \includegraphics[width=7cm]{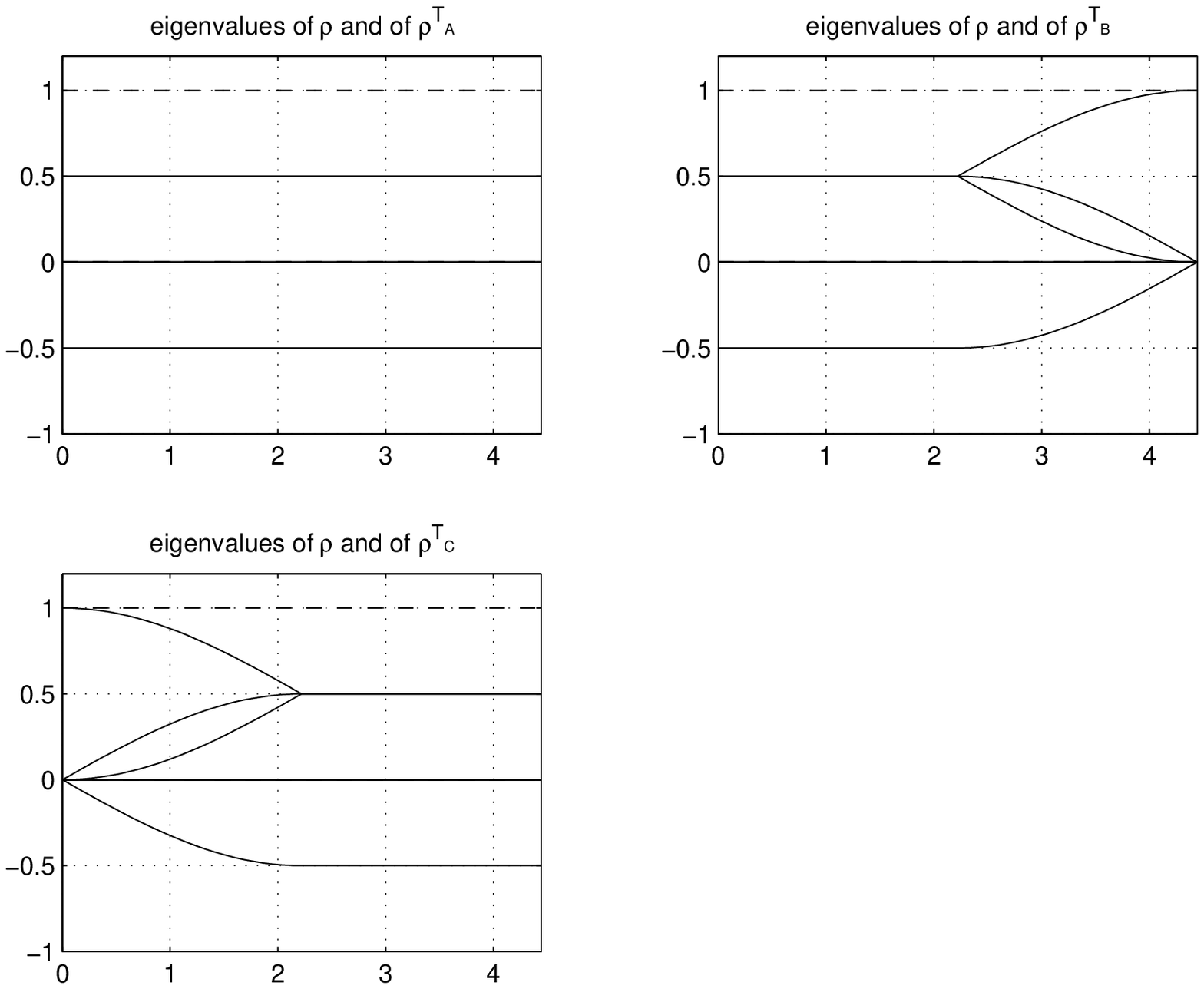}
 \caption{The eigenvalues of $ \rho $ (dashed lines) and of the 3 partial transposes $ \rho^{T_A} $, $ \rho^{T_B} $ and $ \rho^{T_C} $ (solid lines) versus $ t $.}
\label{fig:ex3b}
\end{center}
\end{figure}

Assume we have available coupling Hamiltonians between A and B and between B and C.
B can be thought of as an ancilla being first entangled with A and then sent to interact with C.
Given a state in which A is maximally entangled with B while C is separable from the two (and known), we want to transfer the entanglement from the pair (AB) to the pair (AC) leaving B unentangled at the end of the evolution, without making use of coupling Hamiltonians between A and C.
Assume $  \rho_{AB}(0) $ is in the pure maximally entangled state 
\[
\begin{split}
& \varrho^{\{ 00, \, 11, \, 23, \, 32\}}(0) = \frac{1}{2} \\
& \varrho^{jk}(0) =0 \qquad \text{otherwise} .
\end{split}
\] 
and $ \rho_C = \frac{1}{\sqrt{2}} \left(  \lambda_0 + \lambda_1 \right)$. 
The desired task is accomplished in half of the period $\tau_p =2\sqrt{2} \pi$ for example by the following piecewise constant Hamiltonian: 
\[
-i {\rm ad}_{H} (t) =\begin{cases}
-i {\rm ad}_{\Lambda_{033}} \qquad t \in [0, \, \frac{\tau_p}{4}) \\
-i {\rm ad}_{\Lambda_{220}} \qquad t \in [ \frac{\tau_p}{4}, \, \frac{\tau_p}{2} ].
\end{cases}
\]
We obtain also that $ \rho_{AB}(0) = \rho_{AC}(\pi/2) $ and $ \rho_{B}(\pi/2) = \rho_C(0) $, see Fig~\ref{fig:ex3a}.
As can be seen from Fig.~\ref{fig:ex3b}, at $ \frac{\tau_p}{4} $ the entanglement swaps from the pair AB to the pair AC.
The scheme can be iterated to $n$ qubits.

\subsection{Three-qubit: entangling at distance (II)}
\label{sec:ex:3qub-ent-at-distII}
\begin{figure}[ht]
\begin{center}
 \includegraphics[width=7cm]{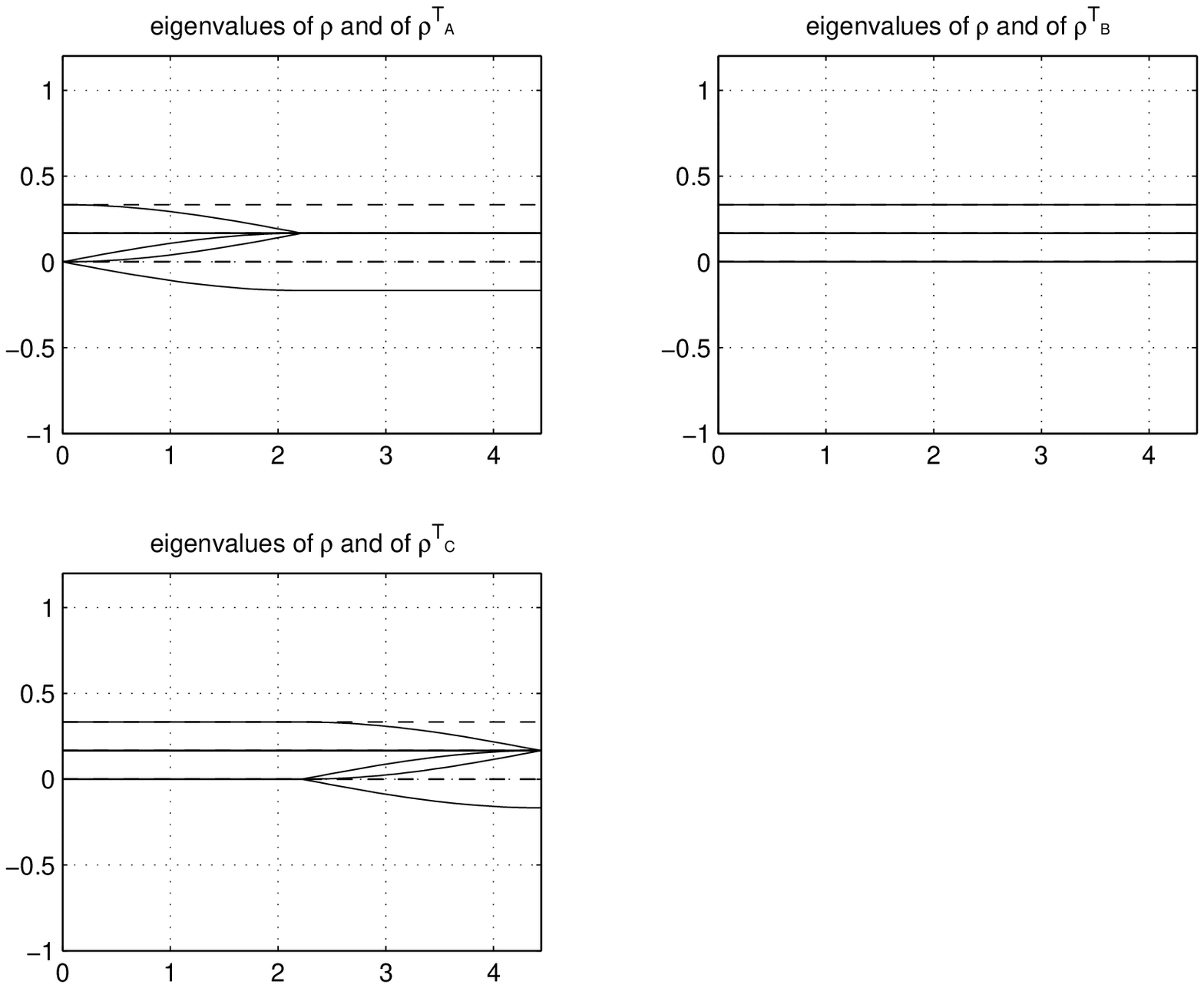}
 \caption{The eigenvalues of $ \rho $ (dashed lines) and of the 3 partial transposes $ \rho^{T_A} $, $ \rho^{T_B} $ and $ \rho^{T_C} $ (solid lines) versus $ t $.}
\label{fig:ex4a}
\end{center}
\end{figure}

\begin{figure}[ht]
\begin{center}
 \includegraphics[width=7cm]{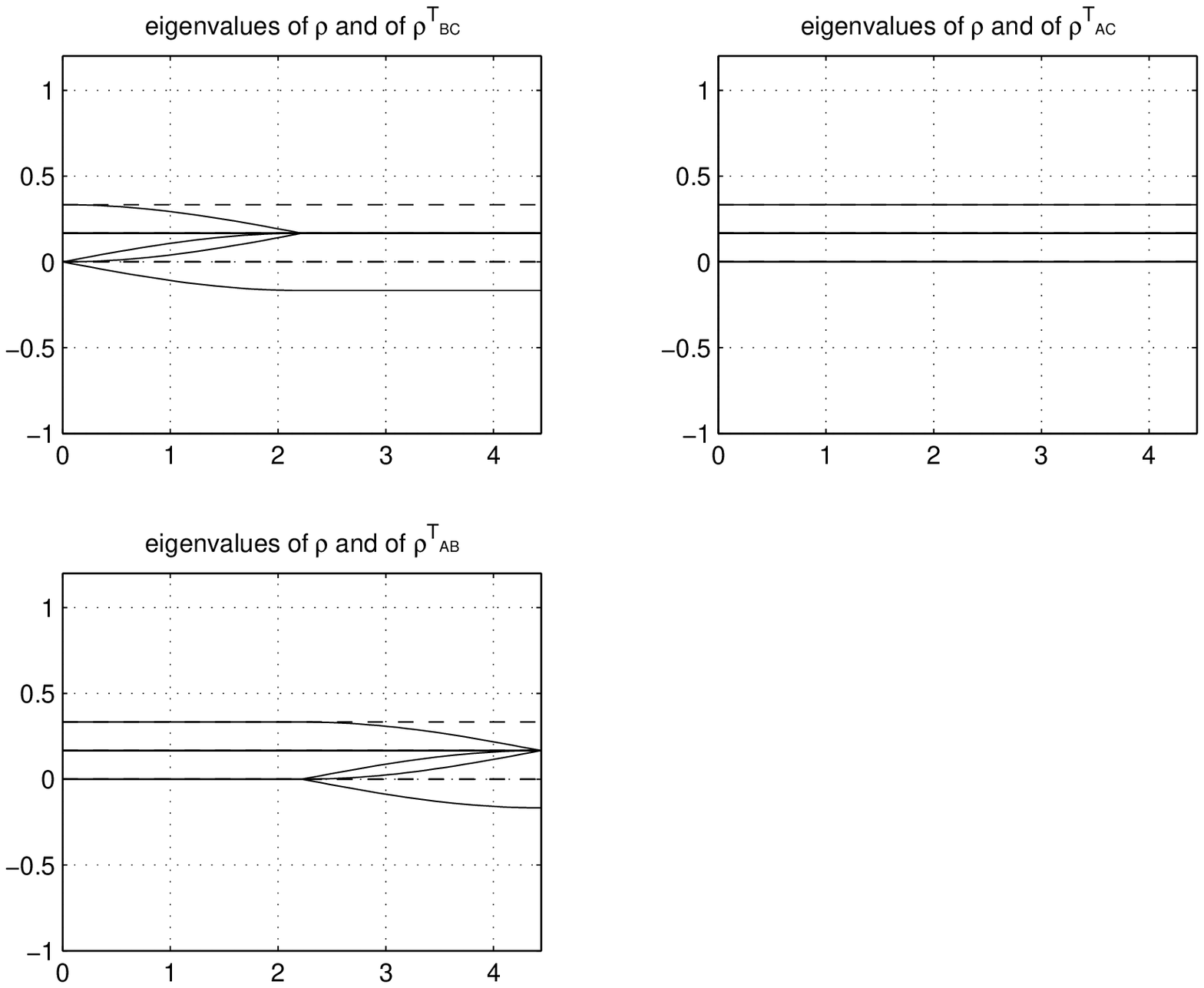}
 \caption{The eigenvalues of $ \rho $ (dashed lines) and of the 3 partial transposes $ \rho^{T_{BC}} $, $ \rho^{T_{AC}} $ and $ \rho^{T_{AB}} $ (solid lines) versus $ t $.}
\label{fig:ex4b}
\end{center}
\end{figure}

While the previous example is rather straightforward, in the literarture there exist more sophisticated and surprising methods to distribute entanglement.
In \cite{Cubitt1} it is shown that for the 3-qubit separable state $ \rho_{\rm in} = \frac{1}{6} \left( \sum_{k=0}^3 \ket{\psi_k, \psi_{-k} , 0} \bra{\psi_k, \psi_{-k} , 0} + \sum_{j=0}^1 \ket{ j, j,1}\bra{ j, j,1} \right) $ with $ \ket{\psi_k } = \left( \ket{0} + e^{ik\pi/2} \ket{1} \right)/\sqrt{2} $, it is possible to find a cascade of two C-NOT gates, one with C as control qubit and acting on A and the other with B as control qubit and acting on C, such that at the end of the operation A and C are both entangled but for the whole process B remains unentangled. 
In terms of the Hamiltonian of the C-NOT computed in Section~\ref{sec:ex:C-NOT}, this is equivalent to the following piece-wise constant 3-qubit infinitesimal generators, obtained permuting the indexes of $ H_{\rm C-NOT} $ and adding a ``0'' in the correct slot \footnote{Notice that the time interval is rescaled with respect to the 2-qubit case of Section~\ref{sec:ex:C-NOT} because of the effect of the third qubit, see \eqref{eq:adj-rep-j0}.}:
\[ \! \! \! \! 
-i {\rm ad}_H(t) =\begin{cases}
-i \left(- {\rm ad}_{\Lambda_{300}} - {\rm ad}_{\Lambda_{001}} + {\rm ad}_{\Lambda_{301}} \right) ,
 \; \;  t \in [0, \, \frac{\pi}{\sqrt{2}}) \\
-i  \left(- {\rm ad}_{\Lambda_{003}} - {\rm ad}_{\Lambda_{010}} + {\rm ad}_{\Lambda_{013}} \right) ,
\; \;  t \in [ \frac{\pi}{\sqrt{2}}, \, \frac{2 \pi}{\sqrt{2}} ].
\end{cases}
\]
If $ x =\frac{1}{6 \sqrt{2} }$, then
\beqan
\rho_{\rm in} & = & \frac{1}{2\sqrt{2} } \Lambda_{000} +x \Lambda_{003} + x \Lambda_{110}+ x \Lambda_{113} \\ 
&& -x \Lambda_{220} -x \Lambda_{223} +x \Lambda_{330}- x \Lambda_{333}, \\
\rho_{\rm int} & = & \frac{1}{2\sqrt{2} } \Lambda_{000} -x \Lambda_{033} + x \Lambda_{111}- x \Lambda_{122} \\
&& -x \Lambda_{212} -x \Lambda_{221} +x \Lambda_{303}+ x \Lambda_{330}, \\
\rho_{\rm fin}  & = & \frac{1}{2\sqrt{2} } \Lambda_{000} -x \Lambda_{030} + x \Lambda_{101}+ x \Lambda_{131} \\ 
&& -x \Lambda_{202} -x \Lambda_{232} +x \Lambda_{303}+ x \Lambda_{333}, 
\eeqan
where $ \rho_{\rm int} $ is the density after the first C-NOT gate and $ \rho_{\rm fin} $ the final state.
Simulating the evolution of the system, we get that indeed B maintains a positive partial transpose (PPT) for the whole interval, as can be seen in Fig.~\ref{fig:ex4a}, while A acquires a negative partial transpose (NPT) in the first half and keeps its through the second half. In this second part also C shows NPT.
The behavior can be explained in terms of bipartite entanglement of different cuts of the 3 qubits.
Compare Fig.~\ref{fig:ex4a} and Fig.~\ref{fig:ex4b} \footnote{Obviously Fig.~\ref{fig:ex4b} is totally redundant with Fig.~\ref{fig:ex4a}: $ \left( \rho^{T_{BC}} \right)^T = \rho^{T_A} $.}.
In the first half of the interval, A is entangling itself with the 2-qubit reduced density $ \rho_{BC} $. Such entanglement is bipartite and is not ``visible'' at the level of 1-qubit reduced densities of B and C. 
The same thing happens between C and (AB) in the second half of the operation.
The example is a well-cooked one as for all times there is no entanglement showing between B and (AC) (not just ``at the end'' of the gate).
The doubt that remains is whether the final result is truly creation of entanlgement between A and C, or rather is only a ``superposition'' of two 1-qubit -- 2-qubit bipartite entanglement.
Notice that a third C-NOT operation on A and C (with either of the two as control qubit) leaves all three qubits with PPT.

\appendix

\section{Formul{\ae} for Lie brackets of tensor product matrices}
\label{app:lie-brack}

\begin{proposition}
Given $ A_1, \ldots, A_n, B_1 , \ldots, B_n  \in M_{m} $, the commutator of $ A_1\ox \ldots \ox A_n $ and $ B_1\ox \ldots \ox B_n $ is given by
\beq
\begin{split}
& [ A_1\ox \ldots \ox A_n , \, B_1 \ox \ldots \ox B_n ]=  \\
& = \sum \frac{1}{2^{n-1}} \left( ( A_1, \, B_1 ) \ox ( A_2, \, B_2 ) \ox \ldots \ox  ( A_n, \, B_n ) \right) 
\end{split}
\label{n-fact-tens-Lieb}
\eeq
where in each summand the bracket $ ( \; \cdot \;, \; \cdot \; ) $ is 
\[
\begin{cases}  [ \; \cdot \;, \; \cdot \; ] \qquad \text{$ k$ times, $ k$ odd} \\
 \{ \; \cdot \;, \; \cdot \; \} \qquad \text{ $ n-k $ times }
\end{cases}
\]
and the sum is over all possible (nonrepeated) combinations of $ [ \; \cdot \;, \; \cdot \; ] $ and $  \{ \; \cdot \;, \; \cdot \; \} $ and over all odd $ k \in [1, \, n] $.

The anticommutator of $ A_1\ox \ldots \ox A_n $ and $ B_1\ox \ldots \ox B_n $ is given by
\beq
\begin{split}
& \{ A_1\ox \ldots \ox A_n , \, B_1 \ox \ldots \ox B_n \} = \\
& = \sum \frac{1}{2^{n-1}} \left( ( A_1, \, B_1 ) \ox ( A_2, \, B_2 ) \ox \ldots \ox  ( A_n, \, B_n ) \right) 
\label{n-fact-tens-Lieantib}
\end{split}
\eeq
where in each summand the bracket $ ( \; \cdot \;, \; \cdot \; ) $ is 
\[
\begin{cases}  [ \; \cdot \;, \; \cdot \; ] \qquad \text{$ k$ times, $ k$ even} \\
 \{ \; \cdot \;, \; \cdot \; \} \qquad \text{ $ n-k $ times }
\end{cases}
\]
and the sum is over all possible (nonrepeated) combinations of $ [ \; \cdot \;, \; \cdot \; ] $ and $  \{ \; \cdot \;, \; \cdot \; \} $ and over all even $ k \in [1, \, n] $.
\end{proposition}
\proof
We will prove the Proposition by induction.
The formula \eqref{n-fact-tens-Lieb} is obviously true for $ n=1 $ (for $ n=2, 3, 4 $ it is explicitly given below).
Assume it is true for $ n-1 $ and write $ \alpha = A_1 \ox \ldots \ox A_{n-1} $, $ \beta = B_1 \ox \ldots \ox B_{n-1} $.
Then for $n$ we have
\[
\begin{split}
 & [ \alpha \ox A_n , \, \beta\ox B_n ]  =  \alpha \beta \ox A_n B_n - \beta \alpha \ox B_n A_n  \\
& \qquad \qquad \pm \frac{1}{2} \left(\alpha \beta\ox B_n A_n +   \beta \alpha \ox A_n B_n \right) 
\\
& = \frac{1}{2} \left(  [ \alpha , \, \beta ] \ox \{ A_n, \,  B_n \} 
+ \{ \alpha, \, \beta \}  \ox [A_n , \, B_n ] \right) .
\end{split}
\]
If $ [ \alpha , \, \beta ] $ contains an odd number of commutators, so does $  [ \alpha , \, \beta ] \ox \{ A_n, \,  B_n \} $.
Likewise, if $  \{ \alpha, \, \beta \} $ has an even number of commutators, $ \{ \alpha, \, \beta \}  \ox [A_n , \, B_n ] $ has to have an odd one.
If $ [ \alpha , \, \beta ] $ and $  \{ \alpha, \, \beta \} $ contain all possible nonrepeated combinations of commutators and anticommutators, so does the expression $ [ \alpha \ox A_n , \, \beta\ox B_n ] $, and the induction is thus completed.
Concerning the anticommutator \eqref{n-fact-tens-Lieantib}, the same induction arguments can be repeated for the following expression:
\[
\begin{split}
& \{ \alpha \ox A_n , \, \beta\ox B_n \}  = \alpha \beta \ox A_n B_n + \beta \alpha \ox B_n A_n \\
& \qquad \qquad \pm\frac{1}{2} \left(\alpha \beta\ox B_n A_n + \beta \alpha \ox A_n B_n \right) 
\\
& = \frac{1}{2} \left(  [ \alpha , \, \beta ] \ox [ A_n, \,  B_n ] 
+ \{ \alpha, \, \beta \}  \ox \{ A_n , \, B_n \} \right) .
\end{split}
\]
\qed
While we are not certain of the complete novelty of the recursive formul{\ae} \eqref{n-fact-tens-Lieb} and \eqref{n-fact-tens-Lieantib}, we are sure that various equivalent variants of them are well-known \footnote{And trivial, since it is enough to replace $  A B = \frac{1}{2} \left( [A, \, B] + \{ A, \, B \} \right)$ in the brute force calculation of the commutator/anticommutator and regroup appropriately.} for low-dimensional tensors.
Restricting to recent related literature, check for example \cite{DAlessandro5,Khaneja2,Havel4}.
The commutators for the first cases used in the paper are given explicitly below.
 
\begin{widetext}
\beq
\begin{split}
[ A_1 \otimes A_2  , \, B_1\otimes B_2 ] &  =  A_1 B_1 \otimes  A_2 B_2 - B_1 A_1\otimes B_2 A_2 
\\
& =  \frac{1}{2} \left(  [ A_1, \, B_1 ] \otimes \{ A_2, \,  B_2 \} +  \{B_1, \,  A_1 \}  \otimes [ A_2, \, B_2 ] \right),
\end{split}
\label{eq:app-Lieb1}
\eeq
\beq
\begin{split}
[ A_1 \otimes A_2\otimes A_3  , \, B_1\otimes B_2\otimes B_3 ] & =   A_1 B_1 \otimes A_2 B_2 \otimes A_3 B_3 - B_1 A_1  \otimes B_2 A_2  \otimes B_3 A_3 \\
& =  \frac{1}{4}  \left( 
[A_1 , \,B_1 ] \otimes \{A_2 , \,B_2 \}\otimes \{ A_3, \, B_3\}
+\{ A_1, \,B_1 \} \otimes [A_2 , \,B_2 ]  \otimes \{A_3 , \,B_3 \}  
 \right. \\ & \left. \quad 
+\{ A_1, \,B_1 \}  \otimes \{A_2 , \,B_2 \}  \otimes [ A_3, \,B_3 ]
+ [A_1, \, B_1] \otimes[A_2 , \,B_2 ]  \otimes [ A_3, \,B_3 ] 
 \right),
\end{split}
\label{eq:app-Lieb2}
\eeq
\beq
\begin{split}
& [A_1 \otimes A_2  \otimes A_3 \otimes A_4, \, B_1 \otimes B_2 \otimes B_3 \otimes B_4]   =
A_1 B_1  \otimes A_2 B_2 \otimes A_3 B_3 \otimes A_4 B_4  
- B_1 A_1 \otimes B_2 A_2 \otimes B_3 A_3 \otimes B_4 A_4  \\
& \quad   =
\frac{1}{8}  ( 
[A_1 , \,B_1 ] \otimes \{A_2 , \,B_2 \}  \otimes \{A_3 , \,B_3 \} \otimes  \{A_4 , \, B_4\} 
+ \{A_1 , \, B_1\}  \otimes [A_2 , \,B_2 ]  \otimes \{ A_3, \,B_3 \} \otimes \{A_4 , \,B_4 \}  
 \\
& \qquad
+ \{A_1 , \,B_1 \} \otimes  \{A_2 , \,B_2 \} \otimes[A_3 , \,B_3 ]  \otimes  \{A_4 , \,B_4 \} 
+ \{A_1 , \,B_1 \} \otimes \{A_2 , \,B_2 \}  \otimes \{ A_3, \,B_3 \} \otimes [A_4 , \,B_4 ]  
\\
& \qquad 
+ [A_1 , \,B_1 ] \otimes [A_2 , \,B_2 ]  \otimes [A_3 , \,B_3 ] \otimes \{A_4 , \, B_4\}  
+ [A_1 , \,B_1 ] \otimes [A_2 , \,B_2 ]  \otimes \{A_3 , \, B_3\} \otimes [A_4 , \, B_4]  
\\
& \qquad
+ [A_1 , \,B_1 ] \otimes \{A_2 , \,B_2 \}  \otimes [A_3 , \,B_3 ] \otimes [A_4 , \,B_4 ]  
+ \{A_1 , \,B_1 \} \otimes [A_2 , \,B_2 ]  \otimes [A_3 , \,B_3 ] \otimes [A_4 , \,B_4 ]  
).
\end{split}
\label{eq:app-Lieb3}
\eeq
\end{widetext}

\bibliographystyle{unsrt}

\end{document}